# WEDGE ABSORBERS FOR MUON COOLING WITH A TEST BEAM AT MICE *

David Neuffer,[#] Fermilab, Batavia, IL 60510 USA,
J. Acosta and D. Summers, University of Mississippi, Oxford, MS 38655 USA
T. Mohayai and P. Snopok, IIT, Chicago, IL 60166 USA

*Abstract*

Emittance exchange mediated by wedge absorbers is required for longitudinal ionization cooling and for final transverse emittance minimization for a muon collider. A wedge absorber within the MICE beam line could serve as a demonstration of the type of emittance exchange needed for 6-D cooling, including the configurations needed for muon colliders. Parameters for this test are explored in simulation and possible experimental configurations with simulated results are presented.

## INTRODUCTION

Ionization cooling has been proposed as a potentially useful method for cooling particle beams.[1, 2] It could be particularly effective in cooling muons, where cooling to emittances suitable for a high-energy lepton collider is possible, and a cooled muon beam in a storage ring could provide an intense source of muon-decay neutrinos (a "neutrino factory"). Ionization cooling could also be used in some proton and ion beam scenarios. In ionization cooling the beam passes through a material, where energy loss is opposite the direction of momentum, followed by longitudinal acceleration restoring the longitudinal momentum This reduces transverse emittances ($\varepsilon_{i,N}$), following the cooling equation:

$$\frac{d\varepsilon_N}{ds} = -\frac{g_t}{\beta^2 E}\frac{dE}{ds}\varepsilon_N + \frac{\beta_\perp E_s^2}{2\beta^3 m_\mu c^2 L_R E}, \quad (1)$$

where the first term is the frictional cooling effect and the second is the multiple scattering heating term. Here $L_R$ is the material radiation length, $\beta_\perp$ is the betatron focusing function, and $E_s$ is the characteristic scattering energy (~14 MeV), and $g_t$ is the transverse partition number.

Longitudinal cooling depends on having the energy loss mechanism such that higher-energy muons lose more energy The equation is:

$$\frac{d\varepsilon_L}{ds} = -\frac{\partial\left(\frac{dE_\mu}{ds}\right)}{E_\mu}\varepsilon_L + \frac{\beta\gamma\,\beta_L}{2}\frac{d\left\langle\left(\frac{\delta p}{p}\right)^2\right\rangle}{ds} \quad (2)$$

The dependence of energy loss with energy is antidamping for $p_\mu < 350$ MeV/c, and only weakly damping for higher energy. However, the longitudinal cooling rate is enhanced when absorbers are placed at non-zero dispersion and the absorber density or thickness depends upon position, such as in a wedge absorber. With wedge cooling, the longitudinal and transverse partition numbers are coupled, exchanging transverse and longitudinal cooling rates:

$$g_L \Rightarrow g_{L,0} + \frac{\eta\rho'}{\rho_0}; g_x \Rightarrow 1 - \frac{\eta\rho'}{\rho_0} \quad (3)$$

where $\rho'/\rho_0$ is the change in density with respect to transverse position, $\rho_0$ is the reference density associated with $dE/ds$, and $\eta$ is the dispersion ($\eta = dx/d(\Delta p/p)$). This coupling is essential to obtaining effective longitudinal cooling and is therefore needed in any multistage system requiring large cooling factors.

High-luminosity $\mu^+$-$\mu^-$ colliders also require a "final cooling" stage in which transverse emittance is reduced, while longitudinal emittance may increase. This may be done by explicit emittance exchange techniques and energy loss in a wedge absorber is a particularly promising one [4,5].

The goal of the MICE experiment is to explore the conditions for ionization cooling and to demonstrate the effectiveness of components of a µ cooling system. Wedge absorber emittance exchange is an essential component of many cooling systems and the MICE demonstration would be greatly enhanced by a wedge demonstration.

Large exchanges can be obtain within a single wedge and readily measured within the limited scope of the MICE experiment. In this note we explore use of single polyethylene wedges to demonstrate the basic principles of emittance exchange within ionization cooling.

## WEDGE EXCHANGE FORMALISM

A transport matrix based formalism can be used to estimate the exchange effects of a single wedge.[6] Figure 1 shows a stylized view of the passage of a beam with dispersion $\eta_0$ through a wedge absorber. The wedge is approximated as an object that changes particle momentum offset $\delta = \Delta p/P_0$ as a function of $x$, and the wedge is shaped such that that change is linear in $x$. (The change in average momentum $P_0$ is ignored in this approximation, as well as energy straggling and multiple scattering.) The rms beam properties entering the wedge are given by the transverse emittance $\varepsilon_0$, betatron amplitude $\beta_0$, dispersion $\eta_0$ and relative momentum width $\delta_0$. (To simplify discussion the beam is focussed to a betatron and dispersion waist at the wedge: $\beta_0'$, $\eta_0' = 0$.) The wedge transforms the $\delta$ of particles depending on position $x$:

$$\frac{\Delta p}{p} = \delta \rightarrow \delta - \frac{2(dp/ds)\tan\frac{\theta}{2}}{P_0}x = \delta - \delta'x$$

$dp/ds$ is the momentum loss rate in the material ($dp/ds = \beta^{-1}dE/ds$). $2x\tan\theta/2 \cong x\tan\theta$ is the wedge thickness at transverse position $x$ (relative to the central orbit at $x$=0), and $\delta' = 2dp/ds\,\tan\theta/2\,/P_0$ indicates the change of $\delta$ with $x$. The dispersion can be represented by a matrix: $\mathbf{M}_\eta = \begin{bmatrix} 1 & \eta_0 \\ 0 & 1 \end{bmatrix}$, since $x \Rightarrow x + \eta_0 \delta$. The wedge is represented by: $\mathbf{M}_\delta = \begin{bmatrix} 1 & 0 \\ -\delta' & 1 \end{bmatrix}$; therefore $\mathbf{M}_{\eta\delta} = \begin{bmatrix} 1 & \eta_0 \\ -\delta' & 1-\delta'\eta_0 \end{bmatrix}$.

---

*Work supported by FRA Associates, LLC under DOE Contract No. DE-AC02-07CH11359.
[#]neuffer@fnal.gov

Writing the $x$-$\delta$ beam distribution as a phase-space ellipse: $g_0 x^2 + b_0 \delta^2 = \sigma_0 \delta_0$, and transforming the ellipse by standard techniques obtains new coefficients $b_1, g_1, \delta_1$. [6] The momentum width is changed to:

$$\delta_1 = \sqrt{g_1 \sigma_0 \delta_0} = \delta_0 \left[ (1-\eta_0\delta')^2 + \frac{\delta'^2 \sigma_0^2}{\delta_0^2} \right]^{1/2}.$$

The bunch length is unchanged. The longitudinal emittance is therefore changed by the factor $\delta_1/\delta_0$. The transverse emittance is changed by the inverse of this factor. The betatron functions ($\beta_1, \eta_1$) are also changed. Wedge parameters can be arranged to obtain large exchange factors in a single wedge.

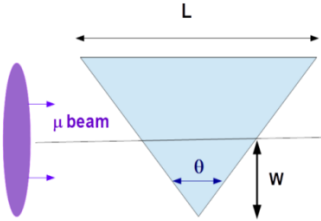

Figure 1: Schematic view of a muon beam passing through a wedge. In dispersive emittance exchange cooling, $\delta g_L = \eta \rho'/\rho_0 = \eta/w$.

## WEDGES FOR FINAL COOLING

In final cooling for a µ collider, we wish to reduce transverse emittance at the cost of increased longitudinal emittance.[3, 4, 5] To obtain a large exchange in a single wedge, the δp spread induced by the wedge should be much greater than the initial value: $\delta_0 \ll \delta'\sigma_0 = \frac{2\tan\left(\frac{\theta}{2}\right)\frac{dp}{ds}}{P_0}\sigma_0$. Thus the incident beam should have a small momentum spread and small momentum $P_0$ and the wedge should have a large $\tan(\theta/2)$, large $dp/ds$ and a large $\sigma_0 = (\varepsilon_0 \beta_0)^{1/2}$. Beam from a final cooling segment is likely to have $P_0 \approx 100$—150 MeV/c, and $\delta p \approx$ 3MeV/c. $\delta p$ should be reduced to ~0.5MeV/c, and this can be done by rf debunching of the beam to a longer bunch length. The best material is a high-density low-Z material (Be or C (diamond density) or $B_4C$ (almost as good)). At these parameters emittance exchange by a factor of 5 to $\varepsilon_x = 25\mu$ can be obtained from a single wedge. This matches the required emittance of a high energy collider, indicating that thick wedge cooling can be an important part of a collider cooling system. Table 1 and figs. 2 and 3 show parameters and ICOOL [9] simulation results of this wedge exchange. (Good agreement with the transport model is also obtained.)

Table 1: Beam parameters at entrance, center and exit of a $w$=3mm, θ=85° diamond wedge. ($z$ = 0, 0.6, 1.2cm) The 0.6cm values can be obtained with a half-size wedge.

| z (cm) | $P_z$(MeV/c) | $\sigma_E$(MeV) | $\varepsilon_x$ (µ) | $\varepsilon_y$(µ) | $\varepsilon_z$ (mm) |
|---|---|---|---|---|---|
| 0 | 100 | 0.5 | 129 | 127 | 1.0 |
| 0.6 | 95.2 | 2.0 | 40.4 | 130 | 4.0 |
| 1.2 | 90.0 | 3.9 | 25.0 | 127 | 7.9 |

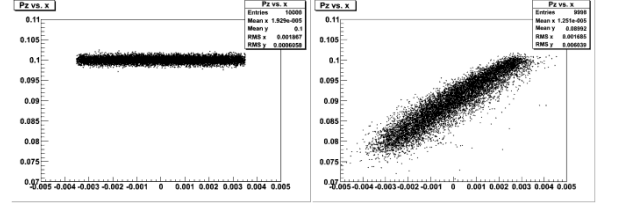

Figure 2: x-P projections of beam before and after the wedge.

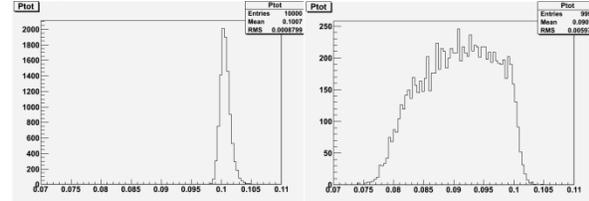

Figure 3: Momentum spread distributions before and after a final cooling wedge.

## EXPERIMENTS AT MICE PARAMETERS

The MICE experiment has considered inserting a wedge absorber into the beam line [7]. The layout would be a scale model of final cooling wedge examples (~10× larger). For an exploration of wedge emittance exchange at present parameters, we would like to include use of 3 Poly wedges: a w=5cm, θ=60° wedge, a w=10cm, θ=30° case and a ~5.5cm thick flat absorber (θ=0°). Poly ($C_2H_4$) is chosen because it is inexpensive, readily available and can be easily machined into the desired shapes. It is a relatively low-Z material (C and H) with relatively little multiple scattering. (Lower Z materials (Li, H, LiH, Be) would be better for cooling, but $C_2H_4$ is adequate for this initial demonstration.)

For the θ=60° wedge, the incident beam would be matched to $\sigma_x$ = 2.5 cm, ($\varepsilon_t$= 3mm, $\beta_t$=36cm) $P_0$=200 MeV/c, corresponding to a baseline MICE beam setting [8], but with $\delta p$ = 2 MeV/c. The small $\delta p$ is obtained by software selection of the incident beam. This example obtains an increase in $\delta p$ by a factor of ~4 accompanied by a reduction in $\varepsilon_x$ by a factor of ~4. This example was simulated in ICOOL [9], with results presented in table 2 and displayed in Fig. 4. The resulting scenario would be an interesting scaled model of a final cooling scenario and would test the basic physics and optics of the exchange configuration. Note that verification of fig. 4 in MICE (and comparison with fig. 3) should be relatively straightforward because it is a large effect, and it would be measuring a critical ingredient in a collider scenario.

Table 2: Beam parameters at entrance, and exit of a w=5 cm, θ=60° C$_2$H$_4$ wedge. (z = 0, 6, 12 cm).

| z (cm) | P$_z$ (MeV/c) | σ$_E$ (MeV) | ε$_x$ (mm) | ε$_y$ (mm) | ε$_z$ (mm) |
|---|---|---|---|---|---|
| 0 | 200 | 1.8 | 3.0 | 3.0 | 2.9 |
| 12 | 182 | 8.6 | 0.76 | 3.0 | 14.3 |

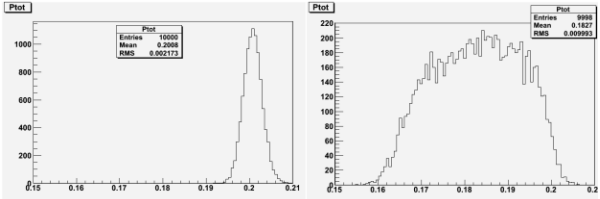

Figure 4: Momentum spread distributions before and after the MICE wedge. (Compare with fig. 3.)

With the 30º absorber, we can explore larger beams, including those with larger δp and dispersion, which can be cooled longitudinally, as well as transverse emittance reduction cases (with small initial δp, similar to the 60º wedge above, but with less exchange). We consider two complementary examples: Case I: a beam with small δp similar to that used for the 60º wedge, and II: a similar transverse emittance beam with large δp and a ~0.5m initial dispersion oriented to reduce longitudinal emittance. The MICE beam has no intrinsic dispersion and a dispersive beam would be generated by software selection of initial beam tracks, much like the small-δp beam.

ICOOL simulations of these simple cases are presented in Table 3. In Case I, emittance exchange is similar to the above 60º case, but the exchange is reduced to a factor of 2.25, following the decrease in angle from 60 to 30. In Case II, the longitudinal emittance is reduced by ~20% while the horizontal emittance increases by a factor of ~30%. Note that, at the parameters presented, the dispersion is matched to nearly zero after the wedge, which may facilitate analysis by decoupling the transverse and longitudinal motion downstream of the absorber. Figure 5 shows x-p distributions before and after the wedge for case II. Verification of this case in MICE would be a first demonstration of longitudinal cooling techniques.

Table 3: Beam parameters at entrance and exit of a w=10 cm, θ=30° C$_2$H$_4$ wedge. (β$_t$=0.5m)

| Case | P$_z$ (MeV/c) | ε$_x$ (mm) | ε$_y$ (mm) | ε$_z$ (mm) | δE$_{,rms}$ MeV | η(m) |
|---|---|---|---|---|---|---|
| I before | 200 | 3.06 | 2.97 | 2.92 | 1.82 | 0.0 |
| I after | 183.3 | 1.36 | 3.08 | 8.18 | 4.76 | 1.14 |
| II before | 200 | 2.95 | 3.03 | 14.5 | 8.82 | 0.51 |
| II after | 183.3 | 3.82 | 3.09 | 12.3 | 7.60 | 0.02 |

With the 0º absorber, no wedge-specific effects occur, and only energy-loss transverse cooling can occur.

However, at expected MICE beam parameters (β$_t$ = 0.5m), the equilibrium emittance is ~5mm with Poly absorbers. With the ~3mm beams presented above, the beam is heated by the absorber. In order to obtain cooling the initial emittance must be significantly greater. With small ε, a measurement is useful for comparison with wedge results, isolating wedge effects from solid absorbers.

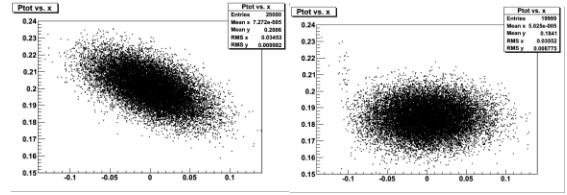

Figure 5: x-p distributions before and after the wedge for Table 3 case II. Note that the wedge has removed most of the dispersion (the x-p correlation).

These ICOOL simulations are based on idealized uncoupled beam passing through the absorbers, and without restricting apertures. In the MICE experiment, solenoidal focusing couples the beam, and x and y emittances are less clearly separated. Some care in emittance reconstruction is required. Simulations within the complete dynamics of MICE, with apertures and magnetic fields, are needed and will specify the degree of agreement with simulation that can be explored in MICE.

## CONCLUSION

Beam transport through wedge absorbers and the resulting changes in beam phase space are a critical ingredient in complete ionization cooling systems; they are essential for longitudinal cooling and important in phase-space manipulations. A demonstration of their effects in MICE would enhance its performance as a first demonstration of the components of ionization cooling systems.